\documentclass[12pt]{article}
\usepackage[xdvi]{graphicx}
\usepackage{epsfig}
\usepackage{amsfonts}
\usepackage{amsbsy}
\usepackage{amsmath}
\usepackage{amssymb}
\usepackage[mathscr]{eucal}

\begin{document}
\newcommand{\sect}[1]{\section{#1}\setcounter{equation}{0}}
\newcommand{\Z}{{\widehat Z}}
\newcommand{\hH}{{\widehat H}}
\newcommand{\enh}{enhan\c con}
\newcommand{\enhs}{enhan\c cons}
\newcommand{\Enh}{Enhan\c con}
\newcommand{\Enhs}{Enhan\c cons}
\newcommand{\re}{{\hat r}_{\rm e}}
\newcommand{\rh}{{\hat r}}
\newcommand{\rL}{r_{\Lambda}}
\newcommand{\beq}{\begin{equation}}
\newcommand{\eeq}{\end{equation}}
\newcommand{\beqq}{\begin{equation*}}
\newcommand{\eeqq}{\end{equation*}}
\newcommand{\beqa}{\begin{eqnarray}}
\newcommand{\eeqa}{\end{eqnarray}}
\newcommand{\reef}[1]{(\ref{#1})}
\newcommand{\labell}[1]{\label{#1}} 
\def\half{\mbox{\scriptsize{${\frac{1}{2}}$}}}
\def\halff{\mbox{\scriptsize{${\frac{5}{2}}$}}}
\def\quarter{\mbox{\scriptsize{${\frac{1}{4}}$}}}
\def\eighth{\mbox{\scriptsize{${\frac{1}{8}}$}}}
\def\bs{\vspace{5pt}}
\def\be{\begin{eqnarray}}
\def\ee{\end{eqnarray}}
\def\ba{\begin{array}}
\def\ea{\end{array}}
\def\bc{\begin{center}}
\def\ec{\end{center}}
\def\bfig{\begin{figure}}
\def\efig{\end{figure}}
\def\ls{\ell_s}
\def\ZZ{\mathbf{Z}}
\def\RR{\mathbf{R}}

\rightline{hep-th/0409280}

\vskip 2cm

\bc

{\bf\Large Generalized Hot \Enhs}

\ec

\vskip 2cm

\renewcommand{\thefootnote}{\fnsymbol{footnote}} \centerline{\bf
David~C.~Page\footnote{page@physics.utoronto.ca},
Amanda~W.~Peet\footnote{peet@physics.utoronto.ca} and
Geoff~Potvin\footnote{gpotvin@physics.utoronto.ca}} \vskip .5cm
\centerline{\it Department of Physics, University of Toronto,}
\centerline{\it Toronto, Ontario, Canada M5S 1A7.}

\setcounter{footnote}{0}
\renewcommand{\thefootnote}{\arabic{footnote}}

\vskip 2cm

\begin{abstract}
We review what has been learnt and what remains unknown about the
physics of hot \enhs \ following studies in supergravity.  We recall a
rather general family of static, spherically symmetric, non-extremal
\enh\ solutions describing D4 branes wrapped on K3 and discuss
physical aspects of the solutions. We embed these solutions in the six
dimensional supergravity describing type IIA strings on K3 and
generalize them to have arbitrary charge vector. This allows us to
demonstrate the equivalence with a known family of hot fractional D0
brane solutions, to widen the class of solutions of this second type
and to carry much of the discussion across from the D4 brane
analysis. In particular we argue for the existence of a horizon branch
for these branes.
\end{abstract}

\vskip 2cm

September, 2004.

\newpage
\section{Introduction}

The celebrated AdS/CFT correspondence \cite{juanadscft}\ has led to
significant advances in the study of string theory and strongly
coupled gauge theories. The original analysis concerned maximally
supersymmetric super-Yang Mills (SYM) theories but was soon extended to
more general gauge theories.

Our focus here is on brane systems giving rise to SYM theories at
large $N$ with ${\mathcal{N}}=2$ supersymmetry and no hypermultiplets.
For these systems, there are strong hints that aspects of the
strong-coupling behaviour of the SYM theories can be understood from
supergravity, despite the lack of a strong/weak duality in the
decoupling limit. This was first shown for the \enh\ system
\cite{Enhancon}.

Going to finite temperature can yield important new information about
the nature of dualities obtained via the decoupling limit from systems
of branes. For the $\mathcal{N}=4$ $SU(N)$ SYM theory in four
dimensions, this was demonstrated in \cite{Witten}\ where various
aspects of the finite temperature gauge theory at large $N$ were found
to be reproduced by the supergravity dual.

Studies of finite temperature \enh\ systems exist in the literature
\cite{HotEnhancon,DPPR,HotFractional} and we proceed to review the
physics of these systems in section 2, adding some new remarks.  We
review the evidence for a novel kind of finite temperature phase
transition in this class of theories.

In section 3, we present a general family of supergravity solutions
for wrapped branes in type IIA on K3 with charges constrained such
that \enh\ behaviour can occur. We also tidy up the literature, by
demonstrating the equivalence between solutions representing D4 branes
wrapped on K3 and fractional D0 branes on the $T^4/\mathcal{Z}_2$
orbifold limit of K3. We motivate why black hole uniqueness theorems
likely specify the physics of the general horizon branch completely.
We also broaden the class of shell branch solutions on the fractional
brane side.

In section 4, we perform some explicit translations of hot \enh\
physics into fractional brane language to resolve a puzzle in the
literature. Section 5 contains some open problems and speculations
about future directions.

\section{Observations on previous related work}
The \enh\ system was the first setup in which a supergravity dual of
pure $\mathcal{N}=2$ SYM theory with no hypermultiplets was studied
\cite{Enhancon}. It was constructed by wrapping BPS D-branes on a K3
manifold, and studying the resulting geometry. From the supergravity
point of view, the system exhibited a novel singularity resolution
mechanism. Naively, there appeared to be a naked timelike singularity
in the space transverse to the branes, dubbed the repulson, because a
massive particle would feel a repulsive potential which becomes
infinite in magnitude at a finite radius from the naive position of
the branes.  Probing the background with a wrapped D-brane, however,
showed that the $N$ source D-branes do not, in fact, sit at the
origin. Rather, they expand to form a shell of branes, inside of which
the geometry does not, after all, become singular.

In the original \enh\ case, taking the decoupling limit did not result
in a clean duality, in the sense that the supergravity dual of the
strongly coupled gauge theory is not weakly coupled. Nonetheless, a
strong hint of the gauge dual of the \enh \ mechanism was seen, in
terms of nonperturbative corrections to the moduli space of pure
$\mathcal{N}=2$ gauge theory. (Corrections due to finite-$N$ were not
ascertained in the supergravity picture, which was studied without
loop corrections.)

\subsection{Heating up the \enh\ system}
A natural generalisation was to study \enh\ geometries for which the
system gains energy above the BPS bound. An unusual two-branch
structure was found \cite{Enhancon} \cite{HotEnhancon}. One class of
possible solutions had the appearance of a black hole (or black
brane), and was dubbed the horizon branch, while the other appeared to
have an \enh-like shell surrounding an inner event horizon and was
dubbed the shell branch.  Only the shell branch correctly matches onto
the BPS \enh\ solution in the limit of zero energy above extremality
but, for sufficiently high extra energy, both solutions were seen to
be consistent with the asymptotic charges. The presence of the horizon
branch far from extremality was expected, since there, the system
should look like an uncharged black hole, when the energy is highly
dominant over the charge.  Additionally, for the shell branch, fixing
the asymptotic charges did not specify exactly how the extra energy
distributed itself between the inner horizon and the shell. 

Dimitriadis and Ross did a preliminary search \cite{Ross} for a
classical instability that would provide evidence that the two
branches are connected.  Such an instability, which is fundamentally
different in nature from the Gregory-Laflamme instability, could be
interpreted as signalling a phase transition in the dual gauge
theory. Such instability was not found. Also presented was an entropic
argument that, at high mass, the horizon branch should dominate over
the shell branch in a canonical ensemble. In later work \cite{Ross2}, a
numerical study of perturbations of the non-BPS shell branch was
completed, but still no instability was found. An analytic proof of
non-existence of such instabilities could not be found either, owing
to the non-linearity of the coupled equations. Furthermore,
\cite{Ross2} investigated whether the shell branch might violate a
standard gravitational energy condition. Indeed, they found that the
shell branch violates the weak energy condition (WEC). This matter
will be important for us in a later section, and so we review it here.

In general, the WEC demands that $T_{\mu\nu}v^{\mu}v^{\nu} \ge 0\,,$
where $v^{\mu}$ is any timelike vector.  For static geometries such as
the heated-up \enh, this condition reduces to
\be \rho \ge 0 \,, \quad \rho + P \ge 0 \,, \ee
where $\rho$ is the energy density and $P$ is the pressure.  The shell
branch solution for the hot \enh\ system has $N$ source branes
located, owing to the \enh\ mechanism, at an incision radius $r_i$
rather than at $r=0$. Because of the shell, the supergravity fields
are not differentiable at the incision radius; the Israel jump
conditions produce the required stress tensor of the shell of branes
(and their excitations).  Picking the system of D4-branes wrapped on
K3, for definiteness, the energy density of this system at $r_i$ has
the form
\be \rho &\sim& -\frac{Z_0'}{Z_0} - \frac{Z_4'}{Z_4} +
\frac{8}{r_i}\left( \sqrt{\frac{L}{K}} - 1\right) \,. \ee
In this expression the harmonic functions $Z_0$, $Z_4$ are the usual
ones exterior to D4-branes outside the shell; $Z_0,Z_4$ are just
constant in the interior. Also, the functions $K$ and $L$ parameterize
the non-extremality exterior and interior to the shell,
respectively. $L$ is a constant if the interior is flat space; by
Gauss' law, the other options are to have a dilaton black hole inside
and/or a hot gas. Now, to avoid unnecessarily complicating the
analysis, we will take the interior to be flat space.  It would be
possible to paste in a dilaton black hole instead. The jump conditions
tell us that we will put the least stringent constraints on the shell
branch supergravity solutions by taking flat space inside.  We do this
in what follows.

Surprisingly, when the system is near extremality and the asymptotic
volume of the K3 is large, the first two terms combine into a
dominant, negative, contribution. Thus the shell branch violates the
WEC. It was argued \cite{Ross2} that the shell branch should therefore
be regarded as unphysical. Accordingly, the horizon branch should be
considered the dominant, valid, supergravity solution for non-BPS
\enhs, for the range of parameters admitting it. For the region of
parameter space in which no horizon branch exists, other solutions,
more general than those yet considered, might be valid \cite{Ross2}.

In subsequent work on non-BPS \enhs, involving two of the current
authors, we used simple supergravity techniques to find the most
general solutions with the correct symmetries and asymptotic charges
of the hot \enh \ system \cite{DPPR}. We showed that the only non-BPS
solution with a well-behaved event horizon is the horizon branch.

We also found that there exists a class of solutions that are
generalizations of the shell branch.  An example of such a
generalization is a two-parameter family, dubbed `$\kappa$-shell
solutions', of which the old shell branch is a one-parameter subset.
Part of this family actually obeys the weak energy condition, and is
therefore a candidate for the correct physical solution.  Demanding
that the WEC be satisfied, however, only fixes $\kappa$ to obey an
inequality. Since we no longer had a microscopic description of the
non-BPS geometry, and therefore could not rely upon the supergravity
solution being built solely out of D-branes, we could not use a
D-probe analysis to distinguish which of these solutions is the
correct generalization of the shell branch.

A further few comments on our general solutions are in order here.
The general $D=10$ solution for non-BPS D4-branes wrapped on a K3 (of
volume $V$ at infinity) are:
\be\label{gkappashell} dS_{10}^2 &=& -
\frac{e^{2a-6c}}{e^{\half(X_0+X_4)}}dt^2 + e^{2c+\half(X_0+X_4)}
(dR^2+R^2d\Omega_4^2) + e^{\half(X_0-X_4)} ds_{K3}^2 \,, \nonumber \\
4 \Phi &=& 3 X_0 - X_4 \,, \nonumber \\ F_{(4)} &=& Q_4 \epsilon_{S^4}
\,, \nonumber \\ F_{(8)} &=& q_4 \epsilon_{S^4} \wedge \epsilon_{K3}
\,, \ee
where
\be e^{a} &=& \left(1-\frac{r_H^6}{R^6}\right) \,, \nonumber \\ e^{3c}
&=& \left( 1 + \frac{r_H^3}{R^3} \right)^2\left(\frac{R^3
+r_H^3}{R^3-r_H^3} \right)^{A_1} \,, \nonumber \\ e^{X_0} &=& \left(
\frac{R^3-r_H^3}{R^3+r_H^3}\right)^{-\kappa}\left( \beta -
\frac{q_4^2}{144 r_H^6 (A_1+\kappa+1)^2\beta} \left(
\frac{R^3-r_H^3}{R^3+r_H^3}\right)^{2(A_1+\kappa+1)}\right) \,,
\nonumber \\ e^{X_4} &=& \left(
\frac{R^3-r_H^3}{R^3+r_H^3}\right)^{-\gamma}\left( \alpha -
\frac{Q_4^2}{144 r_H^6 (A_1+\gamma+1)^2\alpha} \left(
\frac{R^3-r_H^3}{R^3+r_H^3}\right)^{2(A_1+\gamma+1)}\right) \,, \ee
where $r_H^3\ge 0$ and asymptotic flatness implies that
\be \alpha &=& \half + \half \sqrt{1 + \frac{Q_4^2}{36 r_H^6 (A_1 +
 \gamma+1)^2}} \,, \nonumber \\ \beta &=& \half - \half \sqrt{1 +
 \frac{q_4^2}{36 r_H^6 (A_1 + \kappa+1)^2}} \,. \ee
$Q_4$ is the D4-brane charge and $q_4$ is the induced D0-brane charge
and is related to the D4-brane charge by $q_4=-V_{\star}Q_4/V$. Notice
that there are four parameters in these solutions: $r_H^3, \kappa,
\gamma, A_1$.  We must determine which ranges of parameters give
physically interesting geometries.

The first condition we demand is that these geometries actually
possess an \enh.  To find \enhs, we can study a wrapped D4-brane
probe, which takes the form (in static gauge)
\be S_{\rm probe} &=& - \int dt\, m(R) \sqrt{-{\mathbb{P}}(g)}
e^{-\Phi} + \mu_4 \int {\mathbb{P}}(C_{(5)}) - \mu_0 \int
{\mathbb{P}}(C_{(1)}) \,.  \ee
where the (local) mass of the probe is
\be m(R) &=& \mu_4 V(R) -\mu_0 \,. \ee
$V(R)=V e^{X_0-X_4}$ is the volume of the K3 at a radius $R$, and the
ratio of the D0- and D4-brane charges of the probe is
$\mu_0/\mu_4=V_{\star}$.  The probe action breaks up, as usual, into
potential and kinetic pieces.  The potential terms fail to cancel,
owing to breaking of supersymmetry.  An \enh\ occurs when the probe
becomes massless, i.e. satisfies
\be e^{X_0-X_4}|_{R_e} &=& \frac{V_{\star}}{V} \,.  \ee
As an aside, we can also probe with an ordinary D0-brane and get the
expected result: the D0-brane can pass right through the \enh\ radius.

In order to simplify the relevant expressions for understanding the
\enh\ condition, let us define the following shorthand,
\be \zeta \equiv (A_1+\gamma+1) \,, \qquad \eta \equiv (A_1+\kappa+1)
\,.  \ee
The volume of the K3 varies with radius, as we come in from infinity.
We find that there are five different cases depending on the values of
$\zeta$ and $\eta$.  In particular:

{\underline{Case I: $\eta > 0$ and $\zeta \ge 0$}}

\noindent Here, the story is particularly simple.  We find that, at
some radius greater than $r_H$, the volume of the K3 always shrinks to
zero, indicating that somewhere outside this radius, the K3 has
reached its stringy volume.  Note that the old ($A_1 = 0 = \kappa =
\gamma$) shell solution \cite{HotEnhancon} falls into this category.

{\underline{Case II: $\eta > 0$ and $\zeta < 0$}}

\noindent This is more complicated.  Here, the K3 volume is a ratio of
functions which both have zeroes at some finite distance outside
$r_H$.  If the denominator wins this competition, the K3
decompactifies at a finite radius rather than developing stringy
volume appropriate to the \enh.  Otherwise, i.e. if the numerator
wins, there will be an enhancon shell: the K3 shrinks down to its
stringy volume.

The condition to get an \enh\ rather than a decompactification is
\be \left(\frac{\beta}{\beta-1}\right)^{1/2\eta} >
\left(\frac{\alpha-1}{\alpha}\right)^{1/2|\zeta|} \,.\ee

{\underline{Case III: $\eta \le 0$ and $\zeta < 0$}}

\noindent The K3 volume always blows up at a finite radius.  None of
the Case III solutions has an \enh and they are all expected to be
unphysical.

{\underline{Case IV: $\eta \le 0$ and $\zeta \ge 0$}}

\noindent In this case, the physics depends on the ratio
$|\zeta/\eta|$.  When this ratio is (strictly) less than unity, the
volume of the K3 shrinks to zero at $r_H$, passing through the stringy
volume just outside this, where the \enh\ lives.  Conversely, when
this ratio is (strictly) greater than unity, the K3 decompactifies at
$r_H$ and so this is not really a shell-branch solution.  There is a
third, special, case when $\eta$ and $\zeta$ are both zero.  For this
geometry, there are significant simplifications, and we find the
surprising fact that the K3 volume does not run at all coming in from
infinity.  Clearly, then, this does not have an \enh\ either.  Now,
since $\zeta=0$ and $\eta=0$, there is only one remaining parameter
which we can choose to be $\kappa$.  In fact, we can show that these
solutions are unphysical regardless of the value of $\kappa$, but the
reason differs depending on $\kappa$.  Either the metric is singular
and the dilaton blows up, or the solution violates the BPS bound.

All physical supergravity solutions must obey the BPS bound.  In our case,
this inequality is
\be \label{BPSBound} M \ge M_{BPS} = \frac{\Omega_4}{16\pi G_6} \left(
|Q_4| - |q_4| \right) \,, \ee
where
\be \label{ADMMass} M &=& \frac{3\Omega_4 r_H^3}{4\pi G_6} \left(
(A_1+1)(\alpha+\beta-\frac{2}{3})
+\gamma\alpha+\kappa\beta-\frac{1}{2}(\kappa+\gamma)\right) \,. \ee
This puts a further constraint on the physically admissible values of
the parameters ($r_H^3,A_1,\gamma,\kappa$).

Another condition that physical \enh\ solutions should obey is that
the WEC be satisfied at the location of the \enh\ shell.  This will
give us another (different) inequality that the parameters must
satisfy.  Note that knowledge of the microscopic description of our
shell could be expected to tie down all four parameters, either
partially or completely.  Now, the general WEC at the shell is a messy
expression; to clarify the physics, let us study a simpler subclass of
this solution space.

To illustrate, let us consider the subclass where we set
$A_1=0=\gamma$.  We will call these the $\kappa$-shell solutions.
They are a two-parameter family of solutions obeying two inequalities
(the BPS bound and the WEC at the shell). For fixed charges and mass
above extremality, we can take $\kappa$ to be the independent
parameter.  The two inequalities restrict the range of $\kappa$.  This
range depends on the mass above extremality; in the BPS limit, the
allowed range expands to include $\kappa=0$, which corresponds to the
known BPS \enh\ solution.  In the non-BPS case, some of the range of
$\kappa$ satisfying the WEC at the shell and the BPS bound might not
be physical either; however, we do not have a microphysical model to
settle this question definitively.

It is straightforward to find an expression for the \enh\ radius of
the $\kappa$-shell solutions:\footnote{We could also rewrite this in
terms of the parameters: $Q_4 = -3 R_4^3$, $ q_4 = -3 R_0^3$, $ r_H^3
= \quarter r_0^3$, in order to put the solution exactly in terms of
the language of previous studies \cite{HotEnhancon}. }

\be R_e^3 &=& 2 (\kappa+1) r_H^3 + \frac{2}{(V-V_{\star})}
\left(V_{\star}\sqrt{{\frac{Q_4^2}{36}} + r_H^6} + V
\sqrt{{\frac{q_4^2}{36}}+r_H^6(\kappa+1)^2} \right)\,,  \ee
It is clear that the size of any \enh\ shell must be larger, for a
given fixed mass, than the size of the black hole on the horizon
branch, because otherwise the second law of thermodynamics would be
violated.

Later, it will be useful to have these solutions in a
Schwarzschild-type coordinate system, rather than an isotropic one,
for the transverse space. Defining
\be R^3 &=& \half ( r^3 - 2 r_H^3 \pm r^{\frac{3}{2}}\sqrt{r^3-4
r_H^3}) \,, \nonumber \\ f(r)&\equiv& 1 - \frac{4 r_H^3}{r^3} \,. \ee
we find that the $\kappa$-shell solutions take the more suggestive
form
\be\label{suggestive} dS_{10}^2 &=& - f(r) e^{-\half(X_0+X_4)} dt^2 +
e^{\half(X_0+X_4)} \left(\frac{dr^2} {f(r)} + r^2 d\Omega_4^2 \right)
+e^{\half(X_0-X_4)}ds_{K3}^2 \,, \nonumber \\ e^{X_4} &\equiv& \alpha
- (\alpha-1)f(r) \,, \nonumber \\ e^{X_0}&\equiv& f^{-\half\kappa}
(\beta- (\beta-1)f^{\kappa+1}) \,. \ee

In order to be confident that these supergravity solutions are valid,
we need to know that the ten-dimensional string-frame geometry has
small curvature (in string units) and small dilaton.  For the
geometries which have \enhs, supergravity is valid all the way in to
the shell.  These conclusions hold unless we were to try to take a
decoupling limit: in that case, the supergravity solution breaks down
over a significant domain of the geometry.  This is the reason why
there is no clean duality between $\mathcal{N}=2$ gauge theory with no
hypermultiplets and this \enh\ geometry.

\subsection{Relationship to fractional branes}
In a related context, the geometry of fractional D$p$-branes was
studied \cite{Fractional}.  Fractional branes can be described as
regular D$(p+2)$-branes wrapped on a vanishing two-cycle inside the
$T^4/\ZZ_2$ orbifold limit of K3.  The dual gauge theory is again
$\mathcal{N}=2$ SYM with no hypermultiplets. Attempting to take the
decoupling limit once again fails to yield a clean strong/weak
duality. This happens in a way directly analogous to the original
\enh\ case.

The authors of \cite{Fractional} found supergravity solutions for
fractional branes in six dimensions using two different methods.
First, they used boundary state technology to produce a consistent
truncation of Type II supergravity coupled to fractional brane
sources; second, they related their consistent truncation to the
heterotic theory via a chain of dualities.  The BPS solutions they
found exhibit repulson-like behaviour and an analogous \enh\
phenomenon occurs.

The natural extension of this work was, again, to consider the systems
when energy is added to take them above the BPS bound. In
\cite{HotFractional}, a consistent six-dimensional truncation ansatz
for fractional Dp-branes in orbifold backgrounds was provided, for
general $p=0,1,2,3$. Solutions corresponding to the geometry of
non-BPS fractional branes were found, in analogy to the non-BPS \enh \
work \cite{HotEnhancon}. After imposition of positivity of ADM mass,
half of the solutions were disposed of. One of the remaining solutions
was discarded because it did not have a BPS limit.

Considering the other branch (which we will call the shell branch, by
obvious analogy), those authors concluded that these geometries will
always have an \enh \ shell at arbitrary mass above extremality.  Thus
they concluded that horizons never form, and that the gauge dual of
this phenomenon is also prevented from occurring.  In other words, the
mass density of these solutions was thought to be bounded such that it
is never high enough to form a black hole.

The construction of fractional brane geometries that exhibit the \enh\
mechanism is expected to be dual (through T-duality of type IIA on K3)
to the original \enh \ geometries \cite{Enhancon} \cite{Fractional}
\cite{HotFractional}.  However, in view of work reviewed in the
previous subsection, the conclusion that horizons never form in the
non-BPS fractional brane geometries is puzzling.

To further probe the apparent discord in the behaviour of these two
dual systems, let us consider the energy density of the shell
solutions for the fractional brane geometries. To do this, we match
the exterior metric of the shell branch with a black hole interior to
the shell, as before. For definiteness, we pick the fractional
D2-brane:
\be ds_+^2 &=& - H(r)^{-\frac{3}{4}} f(r) dt^2 + H(r)^{\frac{1}{4}}
\left( \frac{1}{f(r)}dr^2 + r^2 d\Omega_4^2 \right) \,, \nonumber \\
ds_-^2 &=& - H(r_i)^{-\frac{3}{4}}\frac{f(r_i)}{F(r_i)}F(r) dt^2 +
H(r_i)^{\frac{1}{4}} \left( \frac{1}{F(r)}dr^2 + r^2 d\Omega_4^2
\right) \,. \ee
Then, at the shell, which is at incision radius $r_i$, we get an
energy density
\be \rho &\sim& - \frac{H'}{H} + \frac{8}{r_i}\left(
\sqrt{\frac{F}{f}}-1\right) \,.  \ee
Near the BPS limit, the energy density of this shell branch does {\em
not} have a dominant negative contribution. This is to be contrasted
with the previous study of shell branch solutions in the Type IIA on
K3 theory relevant to the \enh.  In fact, $\rho$ can be positive or
negative for the fractional brane case, depending on how the energy
above extremality localizes itself.

We will show that this apparent discord is actually an artifact.  The
hot fractional brane system exhibits the exact dual behavior to that
of the hot \enh. In particular, we will show that the solutions of
\cite{HotFractional} are related by duality to the hot \enh\ solutions
of \cite{HotEnhancon}. By continuously varying the K3 moduli away from
the orbifold point, we can reach solutions in which the shell branch
solutions once again violate the WEC. In the following sections we pin
down the precise map between the two setups, and resurrect the horizon
branch on the fractional brane side. We will also exhibit the
fractional brane equivalent of the $\kappa$-shell solutions.

In order to do this we first embed the D4 brane \enh\ solutions in the
full six dimensional supergravity describing type IIA string thoery
compactified on K3. We then show how to generate a complete T-duality
orbit of solutions (ie. with arbitrary charges for the six dimensional
black hole compatible with an \enh\ mechanism, representing any
suitable choice of wrapped branes.) We also allow arbitrary values of
the K3 moduli at infinity - in the case of wrapped D4 \enhs\ this is a
slight generalisation in that we can also include flat $B$-fields
along the internal directions of the K3.

In order to embed the non-extremal D4 brane solutions of \cite{DPPR}
in the six dimensional supergravity, we display a simple two charge
truncation which describes the solutions studied in \cite{DPPR}. These
solutions can then be lifted straight across into the larger
supergravity theory. In deriving the truncation, it is convenient to
switch to heterotic variables using the well-known duality between
type IIA on K3 and heterotic strings on $T^4$. This is also convenient
for comparing with the fractional brane solutions of
\cite{HotFractional} since that paper presents solutions in the
heterotic frame. However, we should stress that we are performing
T-dualities between different IIA solutions and in principle we could
have worked in IIA variables throughout.

\section{Six-dimensional supergravity}

\subsection{Formalism}
The massless fields of heterotic string theory compactified on a
four-torus (or type IIA on $K3$) are the metric $g_{\mu \nu}$, the
B-field $B_{\mu \nu}$, 24 $U(1)$ gauge fields $A_{\mu}^{(a)}$, $(a =1
\ldots 24)$, the dilaton $\phi$ and a matrix of scalar fields $M$
satisfying: \beq \labell{M} M^T = M, \qquad M^T L M = L. \eeq $L$ is a
symmetric matrix which defines an inner product on
$\mathbb{R}^{4,20}$.  The effective action describing the dynamics of
the supergravity fields in six dimensions\footnote{in conventions
standard for the heterotic theory.}  is given by
\beqa \label{6daction} S \sim \int d^6x \sqrt{-G} e^{- 2\phi} \left[R
+4 \partial_{\mu} \phi \partial^{\nu} \phi - \frac{1}{12} H_{\mu \nu
\rho}H^{\mu \nu \rho} \right. \nonumber \\ \left. \qquad \qquad \qquad
- F_{\mu \nu}^{(a)}(LML)_{ab}F^{(b) \mu \nu} + \frac{1}{8}
Tr(\partial_{\mu} ML \partial^{\mu} ML) \right] \, , \eeqa
where
\beqa F_{\mu \nu}^{(a)} &=& \partial_{\mu} A_{\nu}^{(a)} -
\partial_{\nu} A_{\mu}^{(a)} \nonumber \\ H_{\mu \nu \rho} &=&
(\partial_{\mu} B_{\nu \rho} + 2A_{\mu}^{(a)} L_{ab} F^{(b)}_{\nu
\rho}) + \text{cyclic permutations of } \mu, \nu , \rho \, . \eeqa

The equations of motion for $A_{\mu}^{(a)}$ lead to the conserved
electric charges:
\beq \labell{charge} v^{(a)} = \int_{S^5} e^{- 2 \phi} (L M L)_{ab}
\ast F^{(b)} \, . \eeq
In the classical supergravity theory, these charges can take
arbitrary values, but in the quantum theory they are constrained to
lie on a lattice $\Gamma^{4,20} \subset \mathbb{R}^{4,20}$. (In the
heterotic string the 24 quantized charges are carried by fundamental
string states. They are 4 momenta and 4 winding numbers along the
$T^4$ and 16 $U(1)$ charges in the Cartan subalgebra of the 10d gauge
group. In IIA strings on $K3$, the charges label integer homology
classes in the 24 dimensional $H_*(K3,\mathbb{Z})$. Branes wrapped on
cycles carry these charges.)

The effective action \reef{6daction} is invariant under an O(4,20)
symmetry group which acts as
\beq \labell{symmetry} M \rightarrow \Omega M \Omega^T, \, \,
A_{\mu}^{(a)} \rightarrow \Omega_{ab} A_{\mu}^{(b)}, \, \, G_{\mu
\nu} \rightarrow G_{\mu \nu}, \, \, B_{\mu \nu} \rightarrow B_{\mu
\nu}, \, \, \phi \rightarrow \phi \, . \eeq
This extends to a symmetry of the full string theory if it also acts
on the lattice $\Gamma^{4,20}$. With $\Gamma^{4,20}$ fixed, the
discrete subgroup of lattice automorphisms $O(4,20;\mathbb{Z})$ forms
the T-duality group. The action of $\Omega \in O(4,20)$ on $v^{(a)}$
is \beq v^{(a)} \rightarrow (\Omega^T)^{-1}_{a b} v^{(b)} \, . \eeq

The scalar matrix $M$ labels the different vacua of the
theory\footnote{For IIA on $K3$ it describes the K\"ahler and complex
structure moduli of the $K3$ as well as flat $B$-field components in
the internal space. For heterotic compactifications we shall be more
explicit about the relation of $M$ to 10-dimensional quantities in
the following.}. It will be useful to have a geometrical
interpretation of this matrix. First of all, any $M$ of the form
obtained via dimensional reduction from $D=10$ can be written as
\beq M = \Omega^T_0 \Omega_0 \, ,
\eeq
for some $\Omega_0 \in O(4,20)$. The choice of $\Omega_0$ is unique
up to left multiplication by an element of $O(4) \times O(20)$. Thus
choices of $M$ are labeled by points in \beq \frac{O(4,20)}{O(4)
\times O(20)} \,. \eeq This is the space of positive-definite
four-planes in $\mathbb{R}^{4,20}$.

Let us see this correspondence more directly. Planes are in
one-to-one correspondence with the projection operator onto the
plane. Let $P_+$ be the projection operator onto a positive
four-plane. One such projection operator is given by:
\beq P = \frac{1}{2} (1_{24} + L)
\eeq
and all others are related by:
\beq P_+ = \Omega_0^{-1} P \Omega_0 \,,
\eeq
for some $\Omega_0 \in O(4,20)$.  So we find that $P_+$ is related to
$M$ as:
\beq P_+ = \frac{1}{2}(1_{24} + L M) . \eeq
This geometrical language is particularly convenient for expressing
the mass of BPS charged states. The charge of a state is labeled by a
vector $v^{(a)}$ in the lattice $\Gamma^{4,20}$, as above. The BPS
mass depends on the scalars $M$, and is simply the length of the
projection of $v$ onto the four-plane defined by $M$:
\beq m^2 \sim v \cdot P_+ v =
v^T L P_+ v \, .\eeq
Note that this mass formula is invariant under  $O(4,20)$
transformations.

We shall be particularly interested in BPS states which are massive
at generic points in moduli space (generic $M$) but become massless
at special enhan\c{c}on loci. These states correspond to vectors in
the charge lattice $\Gamma^{4,20}$ of negative length:
\beq \labell{cond} v^T L v < 0 \,. \eeq
They become massless when they are orthogonal to the four-plane
defined by $M$
\beq P_+ v = 0 \, .\eeq

\subsection{Generating solutions}
We would like to generate the widest class of static, spherically
symmetric (non-BPS) \enh-like solutions of the six dimensional
supergravity with action \reef{6daction}.  In order to do this, we
should find solutions with arbitrary asymptotic values for the scalar
moduli $M$, and with arbitrary charge vector $v^{(a)}$, subject only
to the condition \reef{cond} which is necessary so that the state can
become massless at special points in moduli space.
We are looking for solutions representing point-like sources, rather
than string-like ones in six dimensions. The restriction of spherical
symmetry therefore rules out the six-dimensional $B_{(2)}$ being
turned on.  The charge vectors we take to be arbitrary.  Where we have
to make an ansatz in the form of our consistent truncation of
supergravity is in taking only two scalar fields to be excited. As
with other systems, we expect that horizon branch (black hole)
solutions will be unique.  

Our main tool for generating solutions will be the $O(4,20)$ symmetry
\reef{symmetry}.  Indeed, given a solution with arbitrary (constant)
asymptotic value for $M$, we can transform it into a solution with
$M=1$ asymptotically by an $O(4,20)$ transformation and so we can
restrict attention to such solutions.

Furthermore, having fixed $M=1$ asymptotically we still have the
freedom to make transformations in the $O(4) \times O(20)$ subgroup of
$O(4,20)$ which fixes the identity matrix. An $O(4)$ rotation can be
used to fix the direction of the component of $v$ in the four-plane
defined by $M=1$, whilst an $O(20)$ rotation can be used to fix the
direction of the component of $v$ orthogonal to the four-plane. After
fixing these directions, we are left with a two parameter family of
possible boundary conditions given by the magnitudes of these two
components of $v$.

It will be helpful at this stage to recall the relation between the
six-dimensional supergravity and the ten-dimensional heterotic
theory. We perform the dimensional reduction using the conventions of
Sen \cite{Sen1}. After compactification, the massless six dimensional
fields are as follows. There are scalar fields $\hat{G}_{ij},
\hat{B}_{ij},{\hat{A}}^I_{i}$ $(i,j=1\ldots4)$ $(I=1\ldots16)$, coming
from the internal components of the metric, $B$-field and $U(1)^{16}$
gauge fields. These are conveniently assembled into the matrix $M$:
 \beq  M = \left(%
\begin{array}{ccc}
  \hat{G}^{-1} & \hat{G}^{-1}\hat{D} - 1_4 & \hat{G}^{-1} \hat{A} \\
  \hat{D}^T \hat{G}^{-1} - 1_4 & \hat{D}^T \hat{G}^{-1} \hat{D} &
  \hat{D}^T \hat{G}^{-1} \hat{A} \\ \hat{A}^T \hat{G}^{-1} & \hat{A}^T
  \hat{G}^{-1} \hat{D} & \hat{A}^T \hat{G}^{-1} \hat{A} + 1_{16} \\
\end{array}%
\right)\,, \eeq where $\hat{G}, \hat{B}$ and $\hat{A}$ are the matrices
with elements $\hat{G}_{ij}, \hat{B}_{ij}$ and $\hat{A}_i^I$
respectively and we have defined $\hat{D} = (\hat{B} + \hat{G} +
\frac{1}{2}\hat{A} \hat{A}^T)$.

There is also a six dimensional dilaton, related to the ten
dimensional one by
\beq\label{dilatonansatz} e^{- 2\phi} = e^{ -2\phi^{(10)}}
\sqrt{\det{\hat{G}}}\,. \eeq
The six dimensional metric $G_{\mu \nu}$ is defined by the relation:
\beq \label{metricansatz} dS^2_{10} = G_{\mu \nu} dx^{\mu} dx^{\nu} +
\hat{G}_{ij} (dz^i + 2A^{(i)}_{\mu} dx^{\mu}) (dz^j + 2A^{(j)}_{\nu}
dx^{\nu}) \, ,\eeq
which also introduces four $U(1)$ gauge fields $A^{(i)}_{\mu}$. The
remaining 20 $U(1)$ gauge fields are given by:
\beq \label{gaugeansatz} A^{(I+8)}_{\mu} = -(\frac{1}{2}
A_{\mu}^{(10)I} - \hat{A}^I_{i} A_{\mu}^{(i)}) \, , \, \,
A_{\mu}^{(i+4)} = \frac{1}{2} B^{(10)}_{i \mu} - \hat{B}_{ij}
A^{(j)}_{\mu} + \frac{1}{2} \hat{A}^I_{j} A^{(I+8)}_{\mu} \, . \eeq
Finally, the six dimensional B-field is given by:
\beq \label{Bansatz}B_{\mu \nu} = B_{\mu \nu}^{(10)} - 4 \hat{B}_{ij}
A_{\mu}^{(i)} A_{\nu}^{(j)} -2(A_{\mu}^{(i)}A_{\mu}^{(i+4)} -
A_{\nu}^{(i)}A_{\mu}^{(i+4)}) \, .\eeq

The charge of a fundamental string state is given by momenta and
winding numbers on $T^4$ and charges under $U(1)^{16}$. We label these
charges by $v=(n_i,w^i,q^I)$. The lattice inner product in terms of
these charges is
\beq v^T L v = 2n_i w^i -q^I q^I \,.\eeq
In other words, the inner product is
\beq L = \left(%
\begin{array}{ccc}
  0 & 1_4 & 0 \\
  1_4 & 0 & 0 \\
  0 & 0 & -1_{16} \\
\end{array}%
\right) \eeq
in this basis.

Following the discussion above, we should look for a two-charge
truncation of the six-dimensional supergravity. A particular choice of
state which has $(v^T L v < 0)$ is given by a fundamental string with
$n_4 = -w_4 = 1$, i.e. one unit of momentum and minus one unit of
winding number along the $z^4$ direction of the torus. This motivates
making a ten dimensional ansatz in which only the fields which couple
to such a state are turned on.

The truncated supergravity arises if we turn off all of the
$U(1)^{16}$ gauge fields of the ten dimensional theory and further
require that three of the compactified dimensions are flat space with
no fields turned on\footnote{This corresponds to smearing the string
along the three remaining torus directions}. We then make an ordinary
$S^1$ reduction on the final compactified direction. Writing this out
explicitly, our reduction ansatz is:
\beqq dS^2_{10} = G_{\mu \nu} dx^{\mu} dx^{\nu} + (dz^2_1 + dz^2_2 +
dz^2_3) + \hat{G}_{44} (dz^4 + 2 A^{(4)}_{\mu} dx^{\mu})^2 \,, \eeqq
\beqq A_{\mu}^{(8)} = \frac{1}{2} B_{4 \mu} \, , \, \,
\, \, B_{\mu \nu} = B^{(10)}_{\mu \nu} - 2(A_{\mu}^{(4)}A_{\nu}^{(8)} -
A_{\nu}^{(4)}A_{\mu}^{(8)})\,, \eeqq
\beq e^{- 2\phi} = e^{ -2\phi^{(10)}} \sqrt{\hat{G}_{44}}
\, .\eeq
The six dimensional field content is thus $(G_{\mu \nu}, B_{\mu \nu},
\phi)$, two gauge fields $A^{(4)}_{\mu}, A^{(8)}_{\mu}$ and a scalar
field $\hat{G}_{44} \equiv e^K$. Substituting into \reef{6daction}
produces the following action for the truncated theory:
\beqa \label{truncated} S \sim \int d^6x \sqrt{-G} e^{- 2\phi} \left[R
+4 \partial_{\mu} \phi \partial^{\nu} \phi - \frac{1}{12} H^2 \qquad
\qquad \qquad \qquad \right. \nonumber \\ \left. \qquad \qquad \qquad
\qquad - (e^K (F^{(4)})^2 + e^{-K} (F^{(8)})^2 ) - \frac{1}{4}
(\partial_{\mu} K \partial^{\mu} K) \right] \, .\eeqa
By construction, any solution of this theory is also a solution of the
full six dimensional supergravity described by the action
\reef{6daction}. Practically, we will truncate further by setting
$B_{\mu \nu} = 0$. This is because we are looking for particle-like
solutions in six dimensions (rather than, for example, string-like
ones).

Next, we describe how to generate $O(4,20)$ families of solutions from
a given solution of the truncated theory \reef{truncated}. First it is
useful to introduce a basis for $\mathbb{R}^{4,20}$ in which the inner
product $L$ is diagonal.  Defining $Q$ to be the orthogonal matrix
\beq Q = \left(%
\begin{array}{ccc}
  \frac{1}{\sqrt{2}}.1_4  & \frac{1}{\sqrt{2}}.1_4 & 0 \\
  -\frac{1}{\sqrt{2}}.1_4 & \frac{1}{\sqrt{2}}.1_4 & 0 \\
  0 & 0 & 1_{16} \\
\end{array}%
\right) \eeq
it is easy to see that the transformation
\beq L \rightarrow QLQ^T \eeq
puts $L$ into the diagonal form $L={\rm diag}(1_{4},-1_{20})$. We
should also transform the matrix $M$ via $M \rightarrow Q M Q^T$ and
the $U(1)$ gauge fields via $A^{(a)} \rightarrow Q_{ab} A^{(b)}$. The
action \reef{6daction} is invariant under this set of transformations

We now introduce a useful notation for embedding solutions of the
truncated theory \reef{truncated} into the theory \reef{6daction}
written in the new basis. Define a 24-entry column vector by
\beq (v^0)^T = \left( v^0_L ,  v^0_R \right) \eeq
where
\beq (v^0_L)^T = \left( 0,0,0,1\right) \qquad \text{and} \qquad
(v^0_R)^T = \left(0,0,0,1,0,0\right) \eeq
are 4- and 20-vectors respectively. A solution of the truncated theory
\reef{truncated} gives rise to a solution of the full six dimensional
theory \reef{6daction} with
\beqq M = 1_{24} + \left( \begin{array}{cc}
  (\cosh{K} - 1) v^0_L v^{0T}_L  & (\sinh{K})v^0_L v^{0T}_R \\
  (\sinh{K})v^0_R v^{0T}_L & (\cosh{K} - 1)v^0_R v^{0T}_R  \\
\end{array}
\right) \, , \eeqq  \beq \label{embedding}F^{(a)} =  \left(
\begin{array}{c}
  (F^L) v^0_L \\
  (F^R) v^0_R \\
\end{array}
\right)^{(a)} \, , \eeq where  \beq F^L = \frac{1}{\sqrt{2}}
(F^{(4)} + F^{(8)}) \qquad , \qquad F^R = \frac{1}{\sqrt{2}}
(F^{(8)} - F^{(4)}) \, . \eeq
We are interested in solutions for which $K \rightarrow 0$
asymptotically so that $M \rightarrow 1$. We can also shift $\phi$ by
a constant if necessary so that $\phi=0$ asymptotically. The $U(1)$
charge of the solution is then computed using \reef{charge} and we
find
\beq v = \left(
\begin{array}{c}
  q_L v^0_L \\
  q_R v^0_R \\
\end{array}
\right) \, ,\eeq
where we have defined
\beq q_L = \int_{S^5\{r=\infty\}} * F^L \, \eeq
and similarly for $q_R$.

It is straightforward to apply $O(4) \times O(20)$ transformations to
these solutions to generate families of solutions with different $v$.
Such transformations have the form
\beq \Omega = \left(
\begin{array}{cc}
  R_4(v_L) & 0 \\
  0 & R_{20}(v_R) \\
\end{array}
\right) \eeq
where $R_4(v_L)$ is a $4 \times 4$ rotation matrix which rotates the
vector $v^0_L$ into an arbitrary unit length 4-vector $v_L$ and,
likewise, $R_{20}(v_R)$ is a $20 \times 20$ rotation which takes the
vector $v^0_R$ into an arbitrary unit 20-vector
$v_R$.\footnote{Different choices of $R_4(v_L)$ and $R_{20}(v_R)$ for
fixed $v_L , v_R$ act identically on the solution \reef{embedding}.}

After applying the symmetry transformation $M \rightarrow \Omega M
\Omega^T$, $F^{(a)} \rightarrow \Omega_{ab} F^{(b)}$, we generate the
solution
\beqq M = 1_{24} + \left(
\begin{array}{cc}
  (\cosh{K} - 1) v_L v^{T}_L  & (\sinh{K})v_L v^{T}_R \\
  (\sinh{K})v_R v^{T}_L & (\cosh{K} - 1)v_R v^{T}_R  \\
\end{array}
\right) \, , \eeqq
\beq \label{solutions}F^{(a)} =  \left(
\begin{array}{c}
  (F^L) v_L \\
  (F^R) v_R \\
\end{array}
\right)^{(a)} \, . \eeq
The charge of this solution is
\beq v =\left(
\begin{array}{c}
  q_L v_L \\
  q_R v_R \\
\end{array}
\right)
 \, .\eeq
The masslessness condition which determines the enhan\c{c}on radius is
\beq 0 = P_+ v = \frac{1}{2} \left[(\cosh K + 1) q_L +
\sinh K q_R\right] L \left(%
\begin{array}{c}
  v_L \\
  {\displaystyle{\left[\frac{\cosh K -1}{\sinh K}\right]v_R }}\\
\end{array}
\right) \, ,\eeq
or, more simply,
\beq (1 + \cosh K) q_L + \sinh K q_R = 0 \, .\eeq

Finally, we can generate further solutions with arbitrary (constant)
asymptotic values for $M$, by acting on the solutions \reef{solutions}
with the remaining symmetry transformations in $O(4,20)/(O(4) \times
O(20))$. These transformations act transitively on the space of
constant asymptotic values for $M$.

\section{Revisiting hot fractional brane physics}
We now return to the explicit solutions of \cite{DPPR} which were
reviewed in section 2. The formalism of the previous section allows us
to rewrite them in a T-duality covariant way. We then transform to the
variables used for the fractional brane solutions and recover and
extend the solutions of \cite{HotFractional}.

We start from the form of the metric (\ref{gkappashell}) discussed in
section 2.  Reducing to six dimensions and then applying
S-duality(\cite{Sen2}) brings us to the following solution written in
the (heterotic) variables of the last section:
\be \label{hetsoln} dS_6^2 &=& -f(r) e^{-(X_0+X_4)}dt^2 + f(r)^{-1}
dr^2 + r^2 d\Omega_4^2 \,, \nonumber \\ \phi_6 &=& -\quarter (X_0+X_4)
\,, \nonumber \\ F^L_{rt} &=& - \frac{1}{\sqrt{2}r^4} (q_4 e^{-2X_0} +
Q_4 e^{-2X_4}) \,, \nonumber \\ F^R_{rt} &=& - \frac{1}{\sqrt{2}r^4}
(-q_4 e^{-2X_0} + Q_4 e^{-2X_4}) \,, \nonumber \\ K&=& X_0-X_4 \,, \ee
A few clarifying comments on the solution-generating process are in
order.  We have previously indicated that $q_4=-V_{\star}Q_4/V$, but
from the point of view of supergravity these two charges are not
related.  That is to say, the supergravity solution we are considering
solves the equations of motion for any value of $Q_4$ and $q_4$ and we
have two independent charges. We only find out about the relation
between the charges by, for example, probing with a D-brane, a stringy
microscopic object. Thus these solutions provide a suitably general,
two-charge family of `seed' solutions for generating the full orbit of
solutions.  After performing $O(4,20)$ transformations, we can restore
the correct quantization condition on the charges by hand.

Now we want to see how the general non-BPS \enh \ solutions look in
the language of the fractional brane constructions. This will also
allow us to confirm that the old solutions of \cite{HotEnhancon} and
of \cite{HotFractional} are, in fact, related by duality. The Appendix
contains tedious details of this calculation. The result of
transforming the $\kappa$-shell solutions to the fractional brane
frame is
\be \label{kappafrac} dS_6^2 &=& - f H^{-\half}dt^2 +
H^{\half}(f^{-1}dr^2 +r^2 d\Omega_4^2) \,, \nonumber \\ e^{\phi_6} &=&
H^{\quarter}\,, \nonumber \\ G_{aa} &=& \frac{\sqrt{H}}{h_1}\,,
\nonumber \\ \sqrt{2}D &=&
\frac{q_2}{q_1}\left(\frac{h_2}{h_1}-1\right) \,, \nonumber \\ C_t &=&
-\frac{q_2}{2r^3}\frac{1}{H} \left( h_1 + h_2 - \frac{q_1}{2q_2
a}\left(\frac{q_2}{q_1}a-1\right)^2e^{X_4}(e^{X_0}I(r)-1)\right) \,,
\nonumber \\ A_t &=& -\frac{q_1}{2r^3} \frac{1}{h_1}
\left(\frac{q_2}{q_1}a+1 - \left(\frac{q_2}{q_1}a-1
\right)e^{X_0}I(r)\right)\,, \ee
where we have used the shorthand
\beq a = -\frac{q_1}{\sqrt{q_2^2 + 2 q_1^2}} \eeq
and defined the functions
\be h_1 &\equiv& \frac{1}{2} \left( \left(\frac{q_2} {q_1}a+1\right)
e^{X_0} - \left(\frac{q_2} {q_1}a-1\right) e^{X_4}\right) \,,
\nonumber \\ h_2 &\equiv& \frac{1}{2} \frac{q_1}{q_2 a} \left(
\left(\frac{q_2} {q_1}a+1\right)e^{X_0} + \left(\frac{q_2}
{q_1}a-1\right) e^{X_4}\right) \,, \nonumber \\ H &\equiv&
\frac{1}{2a^2}h_1^2 - \frac{q_2^2}{2q_1^2}h_2^2 \,, \ee
and
\be \label{I} I(r) &=& -3 r^3 \int{dr \frac{e^{-2X_0}}{r^4}} \,. \ee
The only property of the latter function that we will need here is
that as $\kappa\rightarrow 0$, $I(r)\rightarrow e^{-X_0}$.  The
fractional brane frame constants $a,q_1,q_2$ are related to the
parameters familiar from the \enh\ frame by
\be Q_4 &=& -\frac{q_1}{6}\left(\frac{q_2}{q_1}+\frac{1}{a}\right) \,,
\nonumber \\ q_4 &=& -\frac{q_1}{6} \left(\frac{q_2} {q_1}-
\frac{1}{a} \right)\,.  \ee

Taking the $\kappa\rightarrow 0$ limit gives the hot fractional brane
solutions of \cite{HotFractional}. In other words, the latter are none
other than the first class of solutions found in the hot \enh\ papers
\cite{HotEnhancon}. Explicitly, for $\kappa=0$,
\be dS_6^2&=& -\frac{f dt^2}{\sqrt{H}} + \sqrt{H}
(f^{-1}dr^2+r^2d\Omega_4^2) \,, \nonumber \\ e^{\phi} &=& H^{\quarter}
\,, \nonumber \\ G_{aa} &=& \frac{H^{\half}}{h_1} \,, \quad a =
6,7,8,9 \,, \nonumber \\ D &=&
\frac{q_2}{\sqrt{2}q_1}\left(\frac{h_2}{h_1}-1\right) \,, \nonumber \\
C_t &=& -\frac{q_2h_3}{Hr^3} \,, \nonumber \\ A_t &=&
-\frac{q_1}{h_1r^3} \,, \ee
where
\be H &=& (1+\half\frac{q_2^2}{q_1^2})h_1^2 - \half \frac{q_2^2}
{q_1^2}h_2^2 \,, \nonumber \\ h_1 &=& 1 - \left(\frac{r_1}{r}\right)^3
\,, \nonumber \\ h_2 &=& 1 - \left(\frac{r_2}{r}\right)^3 \,,
\nonumber \\ h_3 &=& \half ( h_1+h_2) \,, \nonumber \\ f(r) &=& 1 -
\left( \frac{r_0}{r}\right)^3 \,, \nonumber \\ r_1^3 &=& \half r_0^3
+\half \frac{\epsilon_1}{\sqrt{q_2^2+2 q_1^2}} \left[ 2q_1^4+
(q_1^2+q_2^2)r_0^6-2\epsilon_2 q_1^2 \Lambda \right]^{\half} \,,
\nonumber \\ r_2^3 &=& \half r_0^3 +\half \frac{\epsilon_1}
{\sqrt{q_2^2}} \left[ 2q_1^4+(q_1^2+q_2^2)r_0^6+2\epsilon_2 q_1^2
\Lambda \right]^{\half} \,, \nonumber \\ \Lambda &=& \left(
q_1^2+(q_1^2+q_2^2) + \quarter r_0^4 \right)^{\half} \,.  \ee
The constants parameterizing non-extremality in the \enh\ frame,
defined by
\beq H_{\rm nonextremal}-1 = \alpha (H_{\rm extremal}-1) \eeq
are related to the fractional brane quantities via
\be \alpha_4^3 &=& \frac{1}{q_1} \left( -\frac{r_1^3}{a} + \frac{q_2
r_2^3}{q_1} \right) \,, \nonumber \\ \alpha_0^3 &=&
\frac{1}{q_1}\left( \frac{r_1^3}{a} + \frac{q_2 r_2^3}{q_1} \right)
\,.  \ee
Our dictionary then tells us that the horizon branch solutions in the
fractional brane language are given by the values $\epsilon_1 =
\epsilon_2 = -1$, while the shell branch solution corresponds to
$\epsilon_1 = \epsilon_2 = +1$.

Recovering the BPS solution is straightforward, by using $r_0
\rightarrow 0$ or equivalently $\alpha_0\rightarrow 1,
\alpha_4\rightarrow 1$. This gives \cite{Fractional},
\be dS_6^2&=& -\frac{dt^2}{\sqrt{H_{\rm bps}}} + \sqrt{H_{\rm bps}}
(dr^2+r^2d\Omega_4^2) \,, \nonumber\\ e^{\phi_6} &=& H_{\rm
bps}^{\quarter} \,,\nonumber \\ G_{aa} &=& H_{\rm bps}^{\half} \,,
\nonumber \\ D &=& -\frac{q_1}{\sqrt{2}r^3} \,, \nonumber\\ C_t &=&
H_{\rm bps}^{-1} -1 \,, \nonumber \\A_t &=& -\frac{q_1}{r^3} \,. \ee
Since
\be H_{\rm bps} &=& 1 + \frac{q_2}{r^3} - \frac{q_1^2}{2r^6} \,, \ee
the charges of \cite{Fractional} and \cite{Enhancon} are consistently
related as
\be -3 q_2 &=& q_4+Q_4 \,, \nonumber \\ 9 q_1^2 &=& -2 q_4 Q_4 \,. \ee
Correctly, this shows that one of $q_4$ or $Q_4$ must be negative --
as appropriate for our system for which the second charge is induced
from the first.

Again, to be confident that these supergravity solutions are valid, we
need to know that the ten-dimensional string-frame geometry has small
curvature (in string units) and small dilaton.  For the geometries
which have \enhs, supergravity is valid all the way in to the shell.
Still there is no clean duality between $\mathcal{N}=2$ gauge theory
with no hypermultiplets and this fractional brane geometry with an
\enh, because taking the decoupling limit ruins the validity of the
supergravity geometry exterior to the shell.

It is also satisfying to study a wrapped brane probe in these
geometries to see where the \enh\ radius occurs.  By following the
duality map, or directly by looking at a fractional brane probe, one
sees that the relevant quantity to study in the fractional brane
duality frame is the flux through the vanishing 2-cycle,
\be b &=& \int_{\mathcal{C}}{B_{(2)}^{10d}} \,. \ee
This vanishes at the \enh \ radius.  In fact, this expression leads
directly to the condition
\be e^{X_0-X_4}|_{r_e} &=& \frac{V_{\star}}{V} \,, \ee
which is the familiar condition that we found when probing the geometry
in the Type II on K3 frame with a wrapped D4 brane.

\section{Discussion}

In \cite{DPPR}, we and co-authors constructed the most general,
static, finite-temperature extensions of the BPS \enh\ solutions of
six-dimensional supergravity possessing spherical symmetry and only
one running modulus: the volume of the K3 on which the D-branes are
wrapped.  In this paper, we generalized the wrapped D4-brane solutions
of \cite{DPPR} to have arbitrary charge vector, i.e.  arbitrary
combinations of D0, D2, and D4 branes wrapped on various cycles in the
K3. We also allowed arbitrary values for the K3 moduli at asymptotic
infinity. We next showed that a particular subset (previously
discovered by \cite{HotEnhancon}) is equivalent to the hot fractional
brane solutions found by \cite{HotFractional}, and thus we widened
this class of solutions. We argued that there is a two-branch
structure (horizon {\em and} shell solutions) in both cases.

The context of this wider class of solutions provides a natural
explanation as to why the WEC looks different in the case of hot
fractional branes. Namely, that the K3 has been taken to a very
special point in moduli space -- the orbifold limit -- and the mass
of a BPS fractional brane probe is fixed. There is no longer the
freedom to set that mass to be large at infinity, which led to the
dominant contribution to the WEC for the original case of D4-branes
wrapped on the K3.

There remain outstanding questions about the stability of these
various branches of solutions. Some instability must exist, because
the horizon branch solutions, which dominate entropically far above
extremality, do not exist below a critical value of the mass. Other
solutions, i.e. the shell branch solutions (or other exotics), must
take over below that point, and connect properly to the known BPS
solutions in the limit. Thus, there must be some unstable mode(s)
driving the transition between these different states near the
critical mass. It is not clear, however, if such a mode is
represented in the bulk supergravity theory.

Another impediment to further progress is the lack of a microphysical
model of the D-brane and string sources giving rise to these non-BPS
\enh\ solutions.  Supergravity alone is apparently insufficient to
settle a number of questions. In particular, for the regime in which
the shell branch solutions exist but the horizon branch ones do not,
the issue of how much energy (above extremality) gets distributed on
the shell, and how much localizes inside the shell in the form of a
black hole and/or a hot gas, is undetermined without knowledge of the
microphysics.

There are, however, indications that arbitrary distributions of
energy above extremality, between the shell, black hole and hot gas,
do not make sense. For example, if the hot \enh\ shell system is very
near extremality and the above-BPS energy is all put into a black
hole in the interior of the shell, the black hole must be tiny. This
indicates that its Hawking temperature will be high, which would lead
to the wrong equation of state for this nearly-BPS system. This
indicates that some kind of phase transition might occur, in which a
black hole could not form in the interior until it became larger than
some critical size. It would be interesting to know if this physics
could be reflected in the physics of the strongly coupled
${\mathcal{N}}=2$ gauge theory.

Also, as we have seen, supergravity allows a number of additional
parameters, for shell branch solutions with given mass and R-R
charges, whose microphysical role is unclear. If we were to allow more
scalar fields to be turned on, further unfixed parameters in the
solutions might be possible. 

One might be tempted to think that these parameters could be fixed by
considering the shell-branch solutions when their mass gets near to
the critical mass at which the horizon branch first appears.  Then an
argument for protecting the second law of thermodynamics (as in
\cite{johnsonmyers}) might give us some information on them.  However,
as shown in \cite{Ross}, the jump in entropy between the two branches
at this point is discontinuous and so more information beyond
supergravity would be required.

\section*{Acknowledgements}
The authors gratefully acknowledge financial support as follows: DCP
from PREA of Ontario; AWP from NSERC and CIAR of Canada, and the
Alfred P. Sloan Foundation; GP from NSERC and the E.C. Stevens
Foundation.  We also wish to thank K.Hosomichi and O. Saremi for
useful discussions.

\section*{Appendix}

In this appendix we re-express the family of solutions of six
dimensional heterotic supergravity, given in equation \reef{hetsoln},
in the variables of the fractional brane solutions. In order to do
this, we make use of the explicit duality map between a class of
ten-dimensional solutions of the heterotic string on $T^4$ and IIA
solutions on $T^4/\mathcal{Z}_2$ which is described in the appendix of
\cite{Fractional}.

The plan is as follows. First, we choose the vectors $v_L$ and $v_R$
in a particular way so that the lift to a ten-dimensional heterotic
solution has a suitable form and then we apply the duality
transformation of \cite{Fractional} to find a type IIA
solution. Finally we reduce to six dimensions to produce the family of
solutions quoted in the main text \reef{kappafrac}.

So we start from the solution \reef{hetsoln} in terms of which the
fields of six dimensional heterotic supergravity are written as:
\begin{equation} \ba{rclrcl}
F &=& \left( \ba{c} F^L v_L \nonumber \\ F^R v_R \ea \right)\,, & M &=& 1
+ \left( \ba{cc} (\cosh{K} -1) v_L v_L^T & \sinh{K} v_L v_R^T
\nonumber \\ \sinh{K} v_R v_L^T & (\cosh{K}-1) v_R v_R^T \ea \right)\,.
\ea
\end{equation}
The trick is picking $v_L$ and $v_R$ correctly.  For reasons that will
be clear shortly, we choose
\be v_L^T &=& \left( \ba{cccc} 0 & 0 & 0 & 1 \ea \right) \,, \nonumber
\\ v_R^T &=& \left( \ba{ccccccccc} 0 & 0 & 0 & \frac{q_2}{q_1}a & a &
a & 0 & ... & 0 \ea \right) \,, \ee
where we have defined
\be (-3q_1)^2 &=& -2 Q_4 q_4 \,, \nonumber \\ -3 q_2 &=& Q_4 + q_4
\,, \nonumber \\ a &=& -\frac{q_1}{\sqrt{q_2^2 + 2 q_1^2}} \,. \ee
To get into the correct
conventions for the lift to ten dimensions we need to apply the transformation
\be F &\rightarrow& Q F \nonumber \\ M &\rightarrow& Q M Q^T \ee
where
\be Q &=& \frac{1}{\sqrt{2}}\left( \ba{ccc} 1 & 1 & 0 \\ -1 & 1 & 0 \\
0 & 0 & \sqrt{2} \ea \right) \,.  \ee
The result is rather a mess:
\be F &\rightarrow& \frac{1}{\sqrt{2}} \left(0_3, F^L +
\frac{q_2}{q_1}a F^R ,0_3,-F^L + \frac{q_2}{q_1}a F^R, \sqrt{2}aF^R ,
\sqrt{2}aF^R ,0_{14} \right)^{T}\,, \ee
\be M &\rightarrow& \left( \ba{cccccccc} 1_3 & 0 & 0 & 0 & 0 & 0 &
... & 0 \\ 0 & M_1 & 0 & M_2 & M_4 & M_4 & ... & 0 \\ 0 & 0 & 1_3 & 0
& 0 & 0 & ... & 0 \\ 0 & M_2 & 0 & M_3 & M_5 & M_5 & ... & 0 \\ 0 &
M_4 & 0 & M_5 & 1+M_6 & M_6 & ... & 0 \\ 0 & M_4 & 0 & M_5 & M_6 &
1+M_6 & ... & 0 \\ ... & ... & ... & ... & ... & ... & ... & 0 \\ 0 &
0 & 0 & 0 & 0 & 0 & 0 & 1 \ea \right)\,, \ee
where
\be M_1 &=& \left[ \half \left( \frac{q_2}{q_1}a+1\right)
e^{\half(X_0-X_4)} - \half \left( \frac{q_2}{q_1}a-1\right)
e^{\half(X_4-X_0)}\right]^2 \,, \nonumber \\ M_6 &=& \half a^2
\frac{(e^{X_0}-e^{X_4})^2}{e^{X_0+X_4}}\,, \ee
and it can be shown that the other functions $M_i$ are related to
these via the following relations:
\be M_2 &=& - M_6 \,, \nonumber \\ M_4^2 &=& M_1 M_6 \,, \nonumber \\
M_1 M_5 &=& - M_4 (1+M_6) \,, \nonumber \\ M_1 M_3 &=& (1+M_6)^2
\,. \ee
Now, the expressions for $M$ and $F$ uniquely determine the following
scalars and gauge fields:
\be G_{bb} &=& 1 \,, \nonumber \\ G_{99} &=& M_1^{-1} \,, \nonumber \\ a_9^I
&=& -\sqrt{\frac{M_6}{M_1}}\,, \nonumber \\ F^{(1)9}_{rt} &=&
\frac{1}{\sqrt{2}} (F^L_{rt} + \frac{q_2}{q_1}a F^R_{rt}) \,, \nonumber \\
F^{(2)}_{rt9} &=& \frac{1}{\sqrt{2}} (-F^L_{rt} + \frac{q_2}{q_1}a
F^R_{rt})\,, \nonumber \\ F^{(3)I}_{rt} &=& a F^R_{rt}\,, \ee
for $b=$6,7,8 and $I=$1 and 2.

In order to complete the lift to ten dimensions, we need to integrate
the field strengths to form gauge potentials. This introduces the
function $I(r)$ of equation \reef{I}. Now, we can use the duality
transformation \cite{Fractional} to convert to a solution of type IIA
on $T^4/\mathcal{Z}_2$:
\be \phi^{(HE)} &=& -\phi_6 \,, \nonumber \\ g_{\mu\nu}^{(HE)} &=&
e^{-2\phi_6}g_{\mu\nu} \,, \nonumber \\ G_{66}^{(HE)} &=&
\sqrt{\frac{G_{77}G_{88}}{G_{66}G_{99}}} \,, \nonumber \\
G_{77}^{(HE)} &=& \sqrt{\frac{G_{66}G_{88}}{G_{77}G_{99}}} \,,
\nonumber \\ G_{88}^{(HE)} &=&
\sqrt{\frac{G_{66}G_{77}}{G_{88}G_{99}}} \,, \nonumber \\
G_{99}^{(HE)} &=& \sqrt{G_{66}G_{77}G_{88}G_{99}} \,, \nonumber \\
A_{\mu}^{(HE)9} &=& C_{\mu} \,, \nonumber \\ A^{(HE)I}_{\mu} +
A^{(HE)I+1}_{\mu} &=& \sqrt{2}A_{\mu}^{I} \,, \nonumber \\ A^{(HE)I}_9
+ A^{(HE)I+1}_{9} &=& -\sqrt{2} D^{I} \,, \ee
for $I=1..16$ in general.  Looking carefully, we see that there is a
change in conventions that must be applied for consistency with our
notation:
\be A_{M}^{I} &\rightarrow& \sqrt{2}A_{M}^{I} \ee
Applying this, we get the following solution:
\be dS_6^2 &=& - f e^{-\half(X_0+X_4)}dt^2 + e^{\half(X_0+X_4)} (
f^{-1}dr^2 +r^2 d\Omega_4^2 ) \,, \nonumber\\ 4\phi_6 &=&
X_0+X_4\nonumber \\ G_{aa} &=& M_1^{-\half} \,, \nonumber \\ \sqrt{2}
D &=& a \frac{e^{X_0} -
e^{X_4}}{\half\left(\frac{q_2}{q_1}a+1\right)e^{X_0} - \half
\left(\frac{q_2}{q_1}a-1\right)e^{X_4}}\,, \nonumber \\ C_t &=&
\frac{-q_1}{4ar^3} \left( \left( \frac{q_2}{q_1}a + 1 \right)^2
e^{-X_4} - \left( \frac{q_2}{q_1}a - 1 \right)^2 I(r)\right)\,,
\nonumber \\ A_t &=& \frac{-q_1}{r^3} \left( \frac{\frac{q_2}{q_1}a +
1 - \left(\frac{q_2}{q_1}a - 1 \right)
I(r)e^{X_0}}{\left(\frac{q_2}{q_1}a+1\right)
e^{X_0}-\left(\frac{q_2}{q_1}a-1\right) e^{X_4} }\right)\,, \ee
where $a=$6,7,8,9.  Note that we have redefined D by a factor of
$-\sqrt{2}$, another necessary change of convention
c.f. \cite{Fractional}.  We can rewrite this solution in a suggestive
way:
\be dS_6^2 &=& - f H^{-\half}dt^2 + H^{\half}(f^{-1}dr^2 +r^2
d\Omega_4^2)\,, \nonumber \\ e^{\phi_6} &=& H^{\quarter} \,, \nonumber
\\ G_{aa} &=& \frac{\sqrt{H}}{h_1}\,, \nonumber \\ \sqrt{2} D &=&
\frac{q_2}{q_1}\left(\frac{h_2}{h_1}-1\right)\,, \nonumber \\ C_t &=&
-\frac{q_2}{2r^3}\frac{1}{H} \left( h_1 + h_2 - \frac{q_1}{2q_2
a}\left(\frac{q_2}{q_1}a-1\right)^2e^{X_4}(e^{X_0}I(r)-1)\right) \,,
\nonumber \\ A_t &=&
-\frac{q_1}{2r^3}\frac{1}{h_1}\left(\frac{q_2}{q_1}a+1 -
\left(\frac{q_2}{q_1}a-1\right)e^{X_0}I(r)\right) \,, \ee
where we have defined the functions
\be h_1 &\equiv& \frac{1}{2} \left(
\left(\frac{q_2}{q_1}a+1\right)e^{X_0} -
\left(\frac{q_2}{q_1}a-1\right)e^{X_4}\right)\,, \nonumber \\ h_2
&\equiv& \frac{1}{2} \frac{q_1}{q_2 a} \left(
\left(\frac{q_2}{q_1}a+1\right)e^{X_0} +
\left(\frac{q_2}{q_1}a-1\right)e^{X_4}\right) \,, \nonumber \\ H&\equiv&
e^{X_0+X_4} \,, \nonumber \\ &\equiv& \frac{1}{2a^2}h_1^2 -
\frac{q_2^2}{2q_1^2}h_2^2\,. \ee
Working out the rest of the physics is taken up in the body of the
paper.


\end{document}